\documentclass[twocolumn,prd,aps,amsmath,amssymb,nofootinbib]{revtex4-1}
\usepackage{graphics}
\usepackage{graphicx}
\usepackage{dcolumn}
\usepackage{bm}
\usepackage{mathrsfs}
\usepackage{pstricks}
\usepackage{color}
\usepackage{slashed}
\usepackage{amsmath}
\usepackage{epsfig}
\usepackage{amsfonts}
\usepackage{amssymb}
\def\beq{\begin{equation}}
\def\eeq{\end{equation}}
\def\bea{\begin{eqnarray}}
\def\eea{\end{eqnarray}}

\def\Re{\textrm{Re}}

\begin{document}

\title{Disordered $\lambda\varphi^{4}+\rho\varphi^{6}$ Landau-Ginzburg model}

\author{R.~Acosta Diaz}
\email{racosta@cbpf.br}
\affiliation{Centro Brasileiro de Pesquisas F\'{\i}sicas, 22290-180 Rio de Janeiro, RJ, Brazil}

\author{G.~Krein}
\email{gkrein@ift.unesp.br}
\affiliation{Instituto de F\'{i}sica Te\'orica, Universidade Estadual Paulista,\\
Rua Dr. Bento Teobaldo Ferraz, 271, Bloco II, S\~ao Paulo, SP 01140-070, Brazil}
\author{N.~F.~Svaiter}
\email{nfuxsvai@cbpf.br}
\affiliation{Centro Brasileiro de Pesquisas F\'{\i}sicas, 22290-180 Rio de Janeiro, RJ, Brazil}

\author{C.~A.~D.~Zarro}
\email{carlos.zarro@if.ufrj.br}
\affiliation{Instituto de F\'isica, Universidade Federal do Rio de Janeiro, 21941-972 Rio de Janeiro, RJ, Brazil}


\begin{abstract}

We discuss a disordered $\lambda\varphi^{4}+\rho\varphi^{6}$ Landau-Ginzburg model defined in a $d$-dimensional space. First we adopt the standard procedure of averaging the disorder dependent free energy of the model. The dominant contribution to this quantity is represented by a series of the replica partition functions of the system. Next, using the replica symmetry ansatz in the saddle-point equations, we prove that the average free energy represents a system with multiple ground states with different order parameters. For low temperatures we show the presence of metastable equilibrium states for some replica fields for a range of values of the physical parameters. Finally, going beyond the mean-field approximation,  the one-loop renormalization of this model is performed, in the leading order replica partition function.
 
\end{abstract}



\maketitle

\section{Introduction}\label{intro}

The critical behavior of disordered systems has been intensively investigated since the 70's using numerical simulations and analytical methods \cite{binder,livro1,livro2,livro3,livro4,nishimori,DLStein}. Two concepts that are of fundamental importance in statistical disordered systems defined on a spatial lattice are  quenched disorder and frustration. In quenched disordered systems, the disorder has slower dynamical evolution than the other dynamical degrees of freedom and, therefore, they can be considered spatially random. Frustration is related to the fact that due to  competing interactions there are situations where it is not possible to find an equilibrium state for the first neighbor spins \cite{frus,and}. These features are realized in the Edwards-Anderson model for a spin-glass system. The model consists of $N$ Ising spins in a $d$-dimensional lattice with finite range interaction where the exchange bonds are randomly ferromagnetic and anti-ferromagnetic \cite{edwards}.  The spin-glass phase is characterized by the fact that at low temperatures there are domains where the spins become randomly frozen in different directions in space. As it has been emphasized by the literature, disordered systems may have infinitely many local equilibrium states \cite{NewStein}. For instance, in the replica symmetry breaking scenario, the free energy landscape for the infinite-ranged spin-glass has a multivalley structure \cite{sk,RBS4,RBS5}. Many different systems beyond the setting of magnetic materials present 
a spin glass-like behavior with an unusual free energy landscape. A fascinating example comes from the field photonics, in that light presents a glassy behavior when propagation in random nonlinear media where the amplitudes of optical modes play a role similar to that of the spins in a magnetic material \cite{pho}.
In addition, there have been investigations of dynamical and pure static properties in the random temperature Landau-Ginzburg model, with special interest on its spin glass-like behavior \cite{ma1, targus}. 

There exist a large variety of statistical models describing quenched disorder, among which those formulated in the continuum are specially interesting in view of their close relationship with quantum field theory (see, $e.g.$, Ref. \cite{efetov}). In these models,  the disorder fields can be separated into two groups. Those which are random external fields and modify the Gaussian contribution in the replica partition function and those that lead to a spatial variation of the couplings of the model. In any case, the standard procedure is averaging the disorder dependent free energy, using for example the replica method \cite{livro3}. Rather than computing the disorder average of the logarithm of the partition function, the replica method consists in averaging the $k$-th power of the partition function, with the average free-energy being obtained in the limit $k\rightarrow 0$. In the mean-field theory of spin-glasses the concept of replica symmetry breaking was introduced \cite{AT} to solve the problem of negative entropy that afflicts the naive replica method, thereby allowing to describe infinitely many pure thermodynamic states of the system. 

Recently, an alternative analytic calculation was proposed to compute the disorder average free-energy for the random source Landau-Ginzburg model \cite{distributional, distributional2}. The method that we named distributional zeta-function method (DZFM) shares similarities with the conventional replica method. For directed polymers and interfaces in random media, Gaussian models \emph{par excellence}, the DZFM and the conventional replica method give the same result \cite{polymer} for the replica symmetric solution. The DZFM was used also to investigated spontaneous symmetry breaking in non-Gaussian models \cite{zarro,robinson}.

One of the motivations of this paper is to emphasize the differences and similarities between the DZFM and the conventional replica method in the study of disordered systems. Here, we employ the DZFM to explore the free-energy landscape of the $d$-dimensional disordered Landau-Ginzburg $\lambda\varphi^{4}+\rho\varphi^{6}$ model. First we write the dominant contribution to the average free energy as a series of the replica partition functions of the  model. Next,  the structure of the replica space is investigated using the saddle-point equations obtained from each replica field theory. Assuming the replica symmetric ansatz,  we prove that the average free energy represents a system with multiple ground states with different order parameters.  This situation is similar to the free energy of the spin-glass phase obtained in the Sherrington-Kirkpatrick model, in the replica symmetry breaking scenario. Also, for low temperatures we show the presence of metastable equilibrium states for some replica fields in a range of values of the physical parameters. One way to describe the spin glass behavior in the low temperature region in the model is to interpret that some terms of  the series representation for the average free energy describes inhomogeneous domains. At low temperatures, some terms of the series may represent macroscopic regions in space, $\Omega^{(k)}$ with a order parameter $\varphi^{(k)}_{0}$. Finally, we perform the one-loop renormalization of this model.

The organization of this paper is the following. In Sec. \ref{sec:random} we discuss the $d$-dimensional random temperature Landau-Ginzburg model. In Sec. \ref{sec:zeta} we discuss the DZFM.
In Sec. \ref{sec:vacuum} we discuss the structure of the replica space using the saddle-point equations of the model. In Sec. \ref{sec:bubble}, the one-loop renormalization of the disordered model is discussed. Conclusions are given in section \ref{sec:conclusions}. We assume that $\hbar=c=k_{B}=1$.

\section{Random temperature Landau-Ginzburg model}\label{sec:random}

In this work we are interested in studying disordered systems through statistical field theory defined in the continuum.  In classical statistical mechanics of Hamiltonian systems, any state is a probability measure on the phase space. The expectation value of any observable can be obtained from an average constructed with the Gibbs measure

\begin{equation}
d\mu_{\text{Gibbs}}=\frac{1}{Z}e^{-\beta H}d\mu_{\text{Liouville}},
\end{equation}

\noindent where $Z$ the partition function, $H$ is the Hamiltonian, $\beta=1/T$, $T$ is the absolute temperature and $d\mu_{\text{Liouville}}$ is the Liouville measure. The partition function is obtained from a normalization procedure.  For systems described in the continuum with infinitely many degrees of freedom, this framework can be maintained. For instance, Euclidean functional methods, with functional of probability measures, introduced classical probabilistic concepts in quantum field theory. The Euclidean correlation functions, i.e., the Schwinger functions, are the analytic continuation of vacuum expectation values for imaginary time of the Wightman functions \cite{schwinger,nakano, symanzik, symanzik2}. For a scalar field, these $n$-point correlation functions, which are the moments of probability measure, are defined by
\begin{equation}\label{eq:correlationfunction}
\langle\varphi(x_{1})..\varphi(x_{k})\rangle=\frac{1}{Z}\int [d\varphi]\prod_{i=1}^{k}\varphi(x_{i}) \exp\left(-S(\varphi)\right), 
\end{equation}
where $[d\varphi]$ is a formal Lebesgue measure, $i.e.$, a measure in the space of all field configurations and $S(\varphi)$ is the Euclidean action of the system.

Let us assume a $\lambda\varphi^{4}+\rho\varphi^{6}$ scalar model without disorder defined in $\mathbb{R}^{d}$. The partition function of the model is defined as
\begin{equation}
Z=\int_{\partial\Omega} [d\varphi]\,\, \exp\bigl(-H(\varphi)\bigr), \label{a8}
\end{equation}
where the effective Hamiltonian is given by $H(\varphi)=H_{0}(\varphi)+H_{I}(\varphi)$.  The free field effective Hamiltonian $H_{0}(\varphi)$ is given by
\begin{equation}
H_{0}(\varphi)=\int d^{d}x\,\frac{1}{2}\varphi(x) \Bigl(-\triangle + m_{0}^{2}\Bigr)\varphi(x)\label{9}
\end{equation}

\noindent  where $\Delta$ is the Laplacian in $\mathbb{R}^{d}$ and $H_{I}(\varphi)$ is the self-interacting non-Gaussian contribution, defined by

\begin{equation}
H_{I}(\varphi)= \int d^{d}x\,\left(\frac{\lambda_{0}}{4} \,\varphi^{4}(x)+\frac{\rho_{0}}{6}\varphi^{6}(x)\right)\label{10}.
\end{equation}

\noindent In Eq. (\ref{a8}), $[d\varphi]$ is  the formal Lebesgue measure, i.e., a measure in the space of all field configurations, given by $[d\varphi]=\prod_{x} d\varphi(x)$, and $\partial\Omega$ in the functional integral means that the field $\varphi(x)$ satisfies some boundary condition. Periodic boundary conditions can be imposed to preserve translational invariance, replacing $\mathbb{R}^{d}$ by the torus $\mathbb{T}^{d}$. 

In order to generate the correlation functions of the model by functional derivatives, as usual a fictitious source is introduced. Therefore, the generating functional of the correlation functions of the model is 

\begin{equation}
Z(j)=\int_{\partial\Omega} [d\varphi]\,\, \exp\left(-H(\varphi)+  \int d^{d}x \,j(x)\varphi(x)\right). \label{8}
\end{equation}

\noindent The $n$-point correlation functions read

\begin{equation}\label{eq:correlationfunction2}
\langle\varphi(x_{1})..\varphi(x_{k})\rangle=Z^{-1}(j)\frac{\delta^{k} Z(j)}{\delta j(x_{1})...\delta j(x_{k})}|_{j=0}. 
\end{equation}
These moments of the probability measure are the sum of all diagrams with $k$ external legs, including disconnected ones, with exception of the vacuum diagrams. The generating functional of $n$-point connected correlation functions can be obtained defining $W(j)=\ln Z(j)$. The order parameter of the model without disorder $\langle\varphi(x)\rangle$ is given by

\begin{equation}
\langle\varphi(x)\rangle=Z^{-1}(j)\left.\frac{\delta Z(j)}{\delta j(x)}\right|_{j=0}. 
\end{equation}

In the following we are interested in discussing the random temperature $d$-dimensional Landau-Ginzburg model. In the Landau-Ginzburg Hamiltonian, if $\lambda_{0}$ and $m_{0}^{2}$ are regular functions of the temperature, a random contribution $\delta m_{0}^{2}(x)$ added to $m_{0}^{2}$ can be  considered as a local perturbation in the temperature. In this case the Hamiltonian of the  model becomes

\begin{align}\label{eq:hamiltonian}
 H(\varphi,\delta m_{0}^{2})&=\int d^{d}x\left[\frac{1}{2}\varphi(x)\Bigl(-\triangle + m_{0}^{2} - \delta m_{0}^{2}(x)\Bigr)\varphi(x)\right. \nonumber \\
&+ \left.\frac{\lambda_{0}}{4}\varphi^{4}(x) + \frac{\rho_{0}}{6}\varphi^{6}(x)\right].
\end{align}

\noindent The $\varphi^{6}$ contribution in the interaction Hamiltonian must be introduced to obtain an Hamiltonian bounded from below, as necessary to correctly describe the critical properties of the model. Br\'ezin and Dominicis \cite{brezin3}, studying a random field model, showed that new interactions should be considered. This term is related to the tricritical phenomenon \cite{gri,gri2}.

The local minima in the Hamiltonian are the configurations of the scalar field that satisfy the saddle-point equations where the solutions depend on the particular configuration of the random mass --- the terms ``random mass" and ``random temperature"; ``false vacuum" and ``metastable equilibrium state"; ``true vacuum" and ``stable equilibrium state"  are used interchangeably throughout the text.  The existence of a large number of metastable states in many disordered systems and the loss of translational invariance makes the traditional pertubative expansion formalism quite problematic. As discussed in the literature, averaging the free energy over the disorder field allows us to implement a perturbative approach in a straightforward way.
 
Let us briefly discuss the $n$-point correlation function associated with a disordered system. The disorder generating functional for one realization of the disorder is given by

\begin{align} \nonumber
 & Z(\delta m^{2}_{0};j)=\\
 & \int_{\partial\Omega} [d\varphi]\,\, \exp\left(-H(\varphi,\delta m_{0}^{2})+\int d^{d}x\, j(x)\varphi(x)\right),
\label{eq:generatingdisorder}
\end{align}

\noindent where a fictitious source, $j(x)$, is introduced. The $n$-point correlation function for one realization of disorder reads

\begin{align}
\left\langle \varphi(x_{1})\right.\cdots&\left.\varphi(x_{n}) \right\rangle_{\delta m_{0}^{2}}=\nonumber \\
&\frac{1}{Z(\delta m_{0}^{2})}\int [d\varphi]\,\prod_{i=1}^{n}\varphi(x_{i})\exp\left(-H(\varphi,\delta m_{0}^{2})\right),
\end{align}

\noindent where the disordered functional integral $Z(\delta m^{2}_{0})=Z(\delta m^{2}_{0},j)|_{j=0}$. 
As in the pure system, one can define a generating functional for one disorder realization, $W_{1}(\delta m^{2}_{0};j)=\ln Z(\delta m^{2}_{0};j)$. Now, we can define a disorder-averaged correlation function as following

\begin{align}
\mathbb{E}&\left[\left\langle \varphi(x_{1})\cdots\varphi(x_{n}) \right\rangle_{\delta m_{0}^{2}}\right]=\nonumber \\
&\int [d \delta m_{0}^{2}] P\left(\delta m_{0}^{2}\right)\left\langle \varphi(x_{1})\cdots\varphi(x_{n}) \right\rangle_{\delta m_{0}^{2}},
\end{align}

\noindent where $\mathbb{E}[\cdots]$ means the average over the ensemble of all the realizations of the quenched disorder, $[d \delta m_{0}^{2}]$ is the formal Lebesgue measure and the probability distribution of the disorder is written as $[d \delta m_{0}^{2}]P\left(\delta m_{0}^{2}\right)$ 
where $P\left(\delta m_{0}^{2}\right)$ is given by

\begin{equation}\label{eq:distprob}
P(\delta m_{0}^{2})=p_{0}\exp\left(-\frac{1}{4\sigma}\int d^{d}x \left(\delta m_{0}^{2}(x)\right)^{2}\right).
\end{equation} 
The quantity $\sigma$ is a small parameter that describes the strength of disorder and $p_{0}$ is a normalization constant. In this case we have a delta correlated disorder field, i.e., $\mathbb{E}[{\delta m^{2}_{0}(x)\delta m^{2}_{0}(y)}]=\sigma\delta^{d}(x-y)$. 
A relevant quantity is the disorder-averaged generating functional $W_{2}(j)=\mathbb{E}[W_{1}(\delta m^{2}_{0};j)]$:

\begin{equation}
W_{2}(j)=\int\,[d\delta m^{2}_{0}]P(\delta m^{2}_{0})\ln Z(\delta m^{2}_{0};j).
\label{eq:disorderedfreeenergy}
\end{equation} 

\noindent Taking the functional derivative of $W_{2}(j)$ with respect to $j(x)$, we get
\begin{equation}
\left.\frac{\delta W_{2}(j)}{\delta j(x)}\right|_{j=0}=\int\,[d\delta m^{2}_{0}]P(\delta m^{2}_{0})\biggl[\frac{1}{Z(h;j)}\left.\frac{\delta Z(h;j)}{\delta j(x)}\biggr]\right|_{j=0}.
\end{equation}

\noindent Since $\langle\varphi(x)\rangle_{\delta m^{2}_{0}}$ is the expectation value of the field for a given configuration of the disorder in the Euclidean field theory with random mass, the above quantity is the averaged normalized expectation value of the field. Taking two functional derivatives of $W_{2}(j)$ with respect to $j(x)$, we get

\begin{align}
\left.\frac{\delta^{2}W_{2}(j)}{\delta j(x_{1})\delta j(x_{2})}\right|_{j(x_{i})=0}&=\mathbb{E}\left[\langle\varphi(x_{1})\varphi(x_{2})\rangle_{\delta m^{2}_{0}}\right]\nonumber \\
& -\mathbb{E}\left[\langle\varphi(x_{1})\rangle_{\delta m^{2}_{0}}\langle\varphi(x_{2})\rangle_{\delta m^{2}_{0}}\right].
\end{align}

\noindent That is, contrary to the pure system case, one finds that $\mathbb{E}\left[\langle\varphi(x_{1})\rangle_{\delta m^{2}_{0}}\langle\varphi(x_{2})\rangle_{\delta m^{2}_{0}}\right]$ is different from $\mathbb{E}\left[\langle\varphi(x_{1})\rangle_{\delta m^{2}_{0}}\right]\mathbb{E}\left[\langle\varphi(x_{2})\rangle_{\delta m^{2}_{0}}\right]$. This fact shows that the taking functional derivatives of $W(j)$ does not lead to connected correlation functions. Indeed, in the disordered system, there is a multivalley structure spoiling the usual perturbative approach, in which the field is expanded around only one minimum \cite{aharony,Parisi88}. One way to tackle this problem is to use the method of spectral zeta-function, which a global approach, $i.e.$, we do not rely upon only one specific minimum. Our aim is to compute 

\begin{equation}
W_{2}(j)|_{j=0}=\int\,[d\delta m^{2}_{0}]P(\delta m^{2}_{0})\ln Z(\delta m^{2}_{0}).
\label{eq:disorderedfreeenergy2}
\end{equation}

In the next section we use the DZFM to calculate the average free energy of the system.

\section{Distributional zeta-function method}\label{sec:zeta}

For free fields without disorder, the spectral zeta-function is a way for regularizing the determinant of the Laplace operator with some boundary conditions and it can be used to calculate the free energy of the system. Here, we are interested in obtaining the average free-energy which is directly related to
$W_{2}(j)|_{j=0}$, after introducing the temperature, i.e., $F=-\frac{1}{\beta}W_{2}(j)|_{j=0}$. Our aim is to compute the disorder-averaged free energy given by
\begin{equation}
F=-\frac{1}{\beta}\int [d\,\delta m_{0}^{2}]P(\delta m_{0}^{2})\,\ln Z(\delta m_{0}^{2}),
\end{equation}
where again $[d\,\delta m_{0}^{2}]$ is also a formal Lebesgue measure.

Recall that a measure space $(\Omega,\mathcal{W},\eta)$ consists in a set $\Omega$,
a $\sigma$-algebra $\mathcal{W}$ in $\Omega$, and a measure $\eta$ on this
$\sigma$-algebra.
Given a measure space $(\Omega,\mathcal{W},\eta)$
and a measurable $f:\Omega\to(0,\infty)$, we define the associated  generalized $\zeta$-function as
\begin{equation*}
\zeta_{\,\eta,f}(s)=\int_\Omega f(\omega)^{-s}\, d\eta(\omega)
\end{equation*}
for those $s\in\mathbb{C}$ such that  $f^{-s}\in L^1(\eta)$,
where in the above integral $f^{-s}=\exp(-s\log(f))$ is obtained using the
principal branch of the logarithm.
%
This formalism contains some well-know examples of zeta-functions, for instance, the classical Riemann zeta-function \cite{riem,riem2}; the prime zeta-function \cite{landau,fro,primes0,primes};
the families of superzeta-functions \cite{voros} and the spectral zeta-functions \cite{haw}.
The usual approach is to define zeta functions in terms of countable collection of numbers, as for example prime numbers, length of closed paths, etc. Here we use 
the definition of the distributional zeta-function $\Phi(s)$, inspired in the spectral zeta-function, as
\begin{equation}
\Phi(s)=\int [d\delta m_{0}^{2}]P(\delta m_{0}^{2})\,\frac{1}{Z(\delta m^{2}_{0})^{s}},\label{pro1}
\end{equation}
\noindent for $s\in \mathbb{C}$, this function being defined in the region where the above integral converges. 
The average free energy can be written as 
\begin{equation}
F=\frac{1}{\beta}(d/ds)\Phi(s)|_{s=0^{+}}, \,\,\,\,\,\,\,\,\,\, \Re(s) \geq 0, 
\end{equation}
\noindent where $\Phi(s)$ is well defined.  To proceed, we use Euler's integral representation for the Gamma function given by
\begin{equation}
\frac{1}{Z(\delta m_{0}^{2})^{s}}=\frac{1}{\Gamma(s)}
\int_{0}^{\infty}dt\,t^{s-1}e^{-Z(\delta m_{0}^{2})t},\,\,\,\,\, \text{for}\,\,\,\, \Re(s)>0.
\label{me5}
\end{equation}
Although the above Mellin integral converges only for $\Re(s)>0$, as  $Z(\delta m_{0}^{2})>0$, we will show how
to obtain from the above expression a formula for the free energy valid for $\Re(s) \geq 0$.
Substituting Eq. \eqref{me5} in Eq. \eqref{pro1} we get
\begin{equation}
\Phi(s)=\dfrac{1}{\Gamma(s)}\int [d\delta m_{0}^{2}]P(\delta m_{0}^{2})\int_0^\infty dt\,t^{s-1}e^{-Z(\delta m_{0}^{2})t}.
\label{pro1.b}
\end{equation}
We already know that the distributional zeta function $\Phi(s)$ is defined for $\mathrm{Re}(s) \geq 0$.
Now we will use the above expression for computing its derivative at $s=0$
by analytic tools.
We assume at principle the commutativity
of the following operations, disorder average, differentiation, integration if necessary.

To continue, take $a>0$ and write $\Phi=\Phi_1+\Phi_2$ where
\begin{equation}
\Phi_{1}(s)=\frac{1}{\Gamma(s)}\int [d\delta m_{0}^{2}]P(\delta m_{0}^{2})\int_{0}^{a}\,dt\, t^{s-1}e^{-Z(\delta m_{0}^{2})t}
\label{m23}
\end{equation}
and
\begin{equation}
\Phi_{2}(s)=\frac{1}{\Gamma(s)}\int [d\delta m_{0}^{2}]P(\delta m_{0}^{2})\int_{a}^{\infty}\,dt\, t^{s-1}e^{-Z(\delta m_{0}^{2})t},
\label{m24a}
\end{equation}
where $a$ is a dimensionless parameter, whose interpretation will be discussed in the next section. The average free energy can be written as
\begin{equation}
F=\left.\frac{1}{\beta}\frac{d}{ds}\Phi_1(s)\right|_{s=0^{\,+}}+\left.\frac{1}{\beta}\frac{d}{ds}\Phi_2(s)\right|_{s=0}\,.
\end{equation}
Let us define the $k$-moment of the distribution, $\mathbb{E}\left[(Z(\delta m_{0}^{2}))^{k}\right]\equiv \mathbb{E}\left[Z^{k}\right]$, where 
\begin{equation}
\mathbb{E}\left[(Z(\delta m_{0}^{2}))^{k}\right]=
\int [d\delta m_{0}^{2}]P(\delta m_{0}^{2}){(Z(\delta m^{2}_{0}))^{k}}.
\end{equation}
The integral $\Phi_{2}(s)$ defines an analytic function defined in the whole complex plane. The contribution of $\Phi_{1}(s)$ reads
\begin{equation}
\Phi_{1}(s)=\frac{a^s}{\Gamma(s+1)}+\frac{1}{\Gamma(s)}\sum_{k=1}^\infty \frac{(-1)^ka^{k+s}}{k!(k+s)}\,
\mathbb{E}\,[{Z^{k}}],
\label{m23c}
\end{equation}
an expression valid for $\Re(s)\geq\,0$.
The function $\Gamma(s)$ has a pole at $s=0$ with residue $1$, therefore
\begin{equation}
\dfrac{d}{ds}\Phi_{1}(s)|_{s=0^{\,+}}=\sum_{k=1}^\infty \frac{(-1)^{k}a^{k}}{k!\,k}\,
\mathbb{E}\,[{Z^{k}}]
+f(a),
\label{m23d}
\end{equation}
where
\begin{equation}
f(a)=\left.\dfrac{d}{ds}\left(\dfrac{a^s}{\Gamma(s+1)}\right)\right|_{s=0}
=\bigl(\log{a}+\gamma\bigr)
\label{m23e2}
\end{equation}
and $\gamma$ is Euler's constant $ 0.577\dots$ The derivative of $\Phi_2$ in Eq. (\ref{m24a}) is given by
\begin{align}
\dfrac{d}{ds}\Phi_{2}(s)|_{s=0}&=\int [d\delta m_{0}^{2}]P(\delta m_{0}^{2})\int_{a}^{\infty}\,\dfrac{dt}{t}\, e^{-Z(\delta m_{0}^{2})t} \nonumber \\
&=R(a). \label{m24}
\end{align}

Hence, using analytic tools, and integrating over the disorder, the average free energy can be represented by

\begin{equation}\label{eq:completefreeenergy}
F=\frac{1}{\beta}\Bigg[\sum_{k=1}^{\infty} \frac{(-1)^{k}a^{k}}{k!\,k}\mathbb{E}\left[Z^{k}\right] +\log{a} + \gamma + R(a)\Bigg],
\end{equation}

\noindent Notice that $R(a)$ vanishes as long as $a\rightarrow\infty$. Indeed, in the following, we discuss the asymptotic behavior of $R(a)$ which is related to the incomplete Gamma function, defined as \cite{abramowitz}

\begin{equation}
\Gamma(\alpha,x)=\int_{x}^{\infty} e^{-t}t^{\alpha-1}\,dt.
\end{equation} 

\noindent The asymptotic representation for  $|x|\rightarrow\infty$ and $-3\pi/2<\text{arg }x<3\pi/2$ reads

\begin{equation}
\Gamma(\alpha,x)\sim x^{\alpha-1}e^{-x}\Big[1+\frac{\alpha-1}{x}+\frac{(\alpha-1)(\alpha-2)}{x^{2}}+\cdots\Big]
\end{equation} 

\noindent  In conclusion, our method is closed related to the use of spectral zeta-functions for
computing the partition function of different systems in quantum field 
theory.  The $\log a$ contribution is similar to the contribution to the free energy $\zeta(0)\log \mu^{2}$ that appears in the the spectral zeta function method, where $\mu$ is a parameter with mass dimension that one has to introduce in order to perform analytic continuations. 

\section{The glassy-like phase in the disordered model}\label{sec:vacuum}

In this section we will discuss the glassy-like phase in the disordered model. From the series representation of the average free energy we have to calculate the $k$-moment of the distribution $\mathbb{E}\left[Z^{k}\right]$. Using the probability distribution for the disorder and the Hamiltonian of the model, the $k$-moment of the distribution is given by

\begin{equation}\label{eq:ezk0}
\mathbb{E}\left[Z^{k}\right]=\int \prod_{i=1}^{k} \left[ d\varphi_{i}\right] e^{-H_{eff}\left(\varphi_{i}\right)},
\end{equation}

\noindent where the effective Hamiltonian, $H_{eff}(\varphi_{i})$ is written as

\begin{align}\label{eq:effectivehamiltonian}
 H_{eff}(\varphi_{i})&=\int d^{d}x\, \left[\frac{1}{2}\sum_{i=1}^{k}\,\varphi_{i}(x)\Bigl(-\Delta\,+m_{0}^{2}\Bigr)\varphi_{i}(x)\right.\nonumber \\
& +\left.\frac{1}{4}\sum_{i,j=1}^{k}g_{ij}\varphi_{i}^{2}(x)\varphi_{j}^{2}(x)
 +\frac{\rho_{0}}{6}\sum_{i=1}^{k}\varphi_{i}^{6}(x)\right],
\end{align}

\noindent where the replica symmetric coupling constants $g_{ij}$ are given by $g_{ij}=(\lambda_{0}\delta_{ij}-\sigma)$.  The saddle-point equations derived from each 
replica partition function read

\begin{align}\label{eq:spe}
\left(-\triangle+m_{0}^{2}\right)\varphi_{i}(x)&+\lambda_{0}\varphi_{i}^{3}(x)+ \rho_{0}\varphi_{i}^{5}(x) \nonumber \\
&-\sigma\varphi_{i}(x)\sum_{j=1}^{k}\varphi_{j}^{2}(x) =0.
\end{align}
Using the replica symmetric ansatz, $\varphi_{i}(x)=\varphi_{j}(x)$, the above equation becomes

\begin{equation}\label{eq:repsymspe}
\left(-\triangle+m_{0}^{2}\right)\varphi_{i}(x)+\left(\lambda_{0}-k\sigma\right)\varphi^{3}_{i}(x) + \rho_{0}\varphi^{5}_{i}(x) =0.
\end{equation}

In the  replica method, using the simplest possible replica symmetric ansatz in each replica partition function, we obtain the saddle-point equations of systems without disorder. 
The replica symmetry breaking scheme was introduced to take into account the presence of many different local minima in the disorder Hamiltonian of the original model. For instance, a  manifestation of this replica symmetry breaking appears in a ferromagnetic system with random spin bonds. There is a low-temperature regime with frustrated spin domains nucleated in a ferromagnetic background \cite{DotHarr}. As we will see, it is possible to obtain a structure with different order parameters in the scenario constructed by the DZFM, where we are not following the standard replica symmetry breaking arguments. Indeed, consider a generic term of the series given by Eq. (\ref{eq:completefreeenergy}) with replica partition function given by $\mathbb{E}\,[{Z^{\,l}}]$ --- see also Eqs. (\ref{eq:ezk0}) and (\ref{eq:effectivehamiltonian}). We are lead to the following choice in the structure of the fields in each replica partition function 

\begin{equation}
\begin{cases}
\varphi_{i}^{(l)}(x)=\varphi^{(l)}(x) \,\,\,\hfill\hbox{for $l=1,2,...,N$}\\
\varphi_{i}^{(l)}(x)=0 \quad \,\,\,\,\,\hbox{for $l>N$},
\end{cases} \label{RSB1}
\end{equation}

\noindent where for the sake of simplicity we still employ the same notation for the field. Therefore the average free energy becomes 

\begin{equation}\label{eq:incompletefreeenergy}
 F=\frac{1}{\beta}\sum_{k=1}^{N} \frac{(-1)^{k}a^{k}}{k!\,k}\mathbb{E}\left[Z^{k}\right]+\cdots.
\end{equation}
In Eq. (\ref{eq:completefreeenergy}), the free energy is independent of $a$. However the entire approach relies on the fact $a$ can be chosen large enough so that $R(a)$ can be neglected in practice. The formalism itself cannot determine a different values of $a$ for which $R(a)$ can be neglected. The specific value must determined phenomenologically.


To proceed, the mean-field theory corresponds to a saddle-point approximation in each replica partition function. A perturbative approach give us the fluctuation corrections to mean-field theory.
Hence, to implement a perturbative scheme, it is necessary to investigate fluctuations around the mean-field equations. Imposing the replica symmetric ansatz
the replica partition function and the effective Hamiltonian for each replica partition function reads
\begin{equation}\label{eq:ezk}
\mathbb{E}\left[Z^{k}\right]=\frac{1}{k!}\int \prod_{i=1}^{k} \left[ d\varphi_{i}^{(k)}\right] e^{-H_{eff}\left(\varphi_{i}^{(k)}\right)},
\end{equation}

\noindent and

\begin{align}\label{eq:effectivehamiltonian2}
 H_{eff}\left(\varphi_{i}^{(k)}\right)&=\int d^{d}x\, \sum_{i=1}^{k}\left[\frac{1}{2}\varphi_{i}^{(k)}(x)\Bigl(-\Delta\,+m_{0}^{2}\Bigr)\varphi_{i}^{(k)}(x)\right. \nonumber \\
 &+\left.\frac{1}{4}\bigl(\lambda_{0}-k\sigma\bigr)\left(\varphi_{i}^{(k)}(x)\right)^{4}+\frac{\rho_{0}}{6}\left(\varphi_{i}^{(k)}(x)\right)^{6}\right].
\end{align}
Note that a $\frac{1}{k!}$ factor was absorbed in $\mathbb{E}\left[Z^{k}\right]$, which can be interpreted to represent an ensemble of $k$-identical replica fields. Also, the fields in each replica partition function are different since each field has the quartic coefficient $(\lambda_{0}-k\sigma)$. Up to now, we have followed the approach developed in Refs. \cite{zarro,robinson}, where we have considered only the leading term in the series representation for the averaged free energy. However, in order to access the glassy-like phases that characterize a disordered system, we have to consider the contributions of all terms in the series given by Eq. (\ref{eq:incompletefreeenergy}). In the following we show that each term in the series in Eq. (\ref{eq:incompletefreeenergy}) describes a field theory with different order parameters.  Therefore a single order parameter is insufficient to describe the low temperature phase of the disordered system.

Here we are following the discussion for the tricritical phenomenon presented in the Ref. \cite{drouffe}.
Let us define a critical $k_{c}$ for each temperature given by

\begin{equation}
k_{c} = \left\lfloor\frac{\lambda_{0}(T)}{\sigma}-\frac{4}{\sigma}\sqrt{\frac{m^{2}_{0}(T)\rho_{0}}{3}}\right\rfloor,
\end{equation}   

\noindent where $\left\lfloor x \right\rfloor$ means the integer part of $x$. Note that $k_{c}$ is a function of $\sigma$, $m_{0}$, $\lambda_{0}$ and $\rho_{0}$. For simplicity, we consider the case where $m_{0}^{2}(T)>0$.  Possible functional forms for the squared mass and coupling constant are $m_{0}^{2}(T)=\mu^{2-\gamma}T^{\gamma}$ and  $\lambda_{0}(T)=\mu^{d-4-\alpha}T^{\alpha}$, where $\mu$, an arbitrary parameter, has mass dimensions.  The region in the parameter space for which $k\leq k_{c}$ corresponds to the situation where  metastability is absent, as the replica fields in each replica partition function fluctuate around zero value, stable equilibrium states.  For $k> k_{c}$, the zero value for the replica fields is a metastable equilibrium state. For these replica partition functions there are first order phase transitions.  The existence of domains with different order parameters can be mostly easily understood in an analogy with a dynamical phase transition induced by a deep temperature quenched \cite{goldenfeld}. Specifically, a system initially in a stable high-temperature equilibrium state will develop spatially inhomogeneous domains when quenched to sufficiently low temperatures. The dynamical evolution stops when the system reaches a new equilibrium state. The nature of the inhomogeneities depends on the equilibrium free energy landscape. In the present case, the inhomogeneities appear in the form of bubble nucleation due to the form of replica free energy in Eq. (\ref{eq:effectivehamiltonian2}), which signals first-order phase transitions.

In the series representation for the free energy, each replica partition function is defined by a functional space where the replica fields are different.  As already discussed, the contribution to the free energy that we are interested is $a$-dependent and, therefore, this structure with multiple ground-states with different order parameters depends on $a$, whose specific value depends on the physical system under consideration. This is a quite interesting situation where the structure of the vacuum states is modified by changing this dimensionless parameter. 
We claim that the series representation for the average free energy lends to natural interpretation of describing inhomogeneous systems. The average free energy,  Eq. (\ref{eq:incompletefreeenergy}) can be written as  

\begin{equation}\label{eq:freeenergya1}
 F=F_{1}+\frac{1}{\beta}\sum_{k=k_{c}}^{N} \frac{(-1)^{k}a^{k}}{k}\mathbb{E}\left[Z^{k}\right]+... \, ,
\end{equation}

\noindent where $F_{1}$ is the contribution to the average free energy for replica fields which oscillate around the true vacuum, $i.e.$, $\varphi^{(k)}_{0}=0$, for $k\leq k_{c}$. $\mathbb{E}\left[Z^{k}\right]$ is defined in Eqs. (\ref{eq:ezk}) and (\ref{eq:effectivehamiltonian2}). 


 One interpretation for this series is that  each  term  describes macroscopic homogeneous domains.  Each domain $\Omega^{(k)}$ has at least one order parameter $\varphi^{(k)}_{0}$. An important question is the size of the domains in the model. The size of each domain is characterized by the correlation length $\xi^{(k)}$ that can be estimated, from the renormalized correlation functions. Therefore in the next section we will perform the one-loop renormalization of the model.

\section{One-loop renormalization in the disordered model}\label{sec:bubble}

To proceed we will go beyond the mean-field approximation by implementing the one-loop renormalization in this model. For the sake of simplicity, we consider the
leading replica partition function. That partition function is described by a large-$N$ Euclidean replica field theory \cite{zarro}. Notice that all the calculations can be computed in a generic replica partition function.   The leading replica partition function is written as
\begin{equation}\label{eq:ezN}
\mathbb{E}\left[Z^{N}\right]=\frac{1}{N!}\int \prod_{i=1}^{N} \left[ d\varphi_{i}\right] e^{-H_{eff}(\varphi_{i})},
\end{equation}
\noindent where

\begin{align}\label{eq:effectivehamiltonian2N}
 H_{eff}(\varphi_{i})&=\int d^{d}x\, \sum_{i=1}^{N}\left[\frac{1}{2}\varphi_{i}(x)\Bigl(-\Delta\,+m_{0}^{2}\Bigr)\varphi_{i}(x)\right.\nonumber \\
 &+\left.\frac{1}{4}(\lambda_{0}-N\sigma)\varphi_{i}^{4}(x)+\frac{\rho_{0}}{6}\varphi_{i}^{6}(x)\right],
\end{align}

\noindent where for simplicity, $\varphi_{i}^{(N)}=\varphi_{i}$. Let us define $g_{0}=\lambda_{0}-f_{0}$, where $f_{0}=N\sigma$. We maintain $f_{0}$ fixed while $N\rightarrow\infty$ and $\sigma\rightarrow 0$. Since in the Landau-Ginzburg scenario $\lambda_{0}$ depends on the temperature, $g_{0}$ is not positive definite for sufficiently low temperatures. For simplicity we assume that $m_{0}^{2}$ is a positive quantity. In this situation we have $N$ replicas with  true and false vacua.  Vacuum transitions in this theory with $N$ replicas can be described in the following way. Lowering the temperature each replica field has a false vacuum and two degenerate true vacuum states. The transition from the false vacuum to the true one will nucleates bubbles of the true vacuum  \cite{SColeman1,SColeman2,Flores}.  One way to proceed is to calculate the transition rates in the diluted instanton approximation. This is a standard calculation that can be found in the literature. Instead of this, our goal is to perform the one-loop renormalization of the model.

At this point, let us introduce an external source $J_{i}(x)$ in replica space linearly coupled with each replica. Considering only the leading term in the series representation for the average free energy, and absorbing the dimensionless quantity $a$ in the functional measure, we are able to define the generating functional of all correlation functions for a large-$N$ Euclidean field theory  as $\mathbb{E}\,[Z^{N}(J)]=\mathcal{Z}(J)$. To proceed we are following the Ref. \cite{Itzy}. Accordingly this generating functional of all correlation functions of this Euclidean field theory is given by 

\begin{equation} 
\mathcal{Z}(J)=\frac{1}{N!}\int \prod_{i=1}^{N} \left[ d\varphi_{i}\right] e^{-H_{eff}(\varphi_{i})+\int d^{d}x \sum_{i=1}^{N}J_{i}\varphi_{i}}.
\end{equation}

It is possible to define the generating functional of connected correlation functions $\mathcal W(\mathcal{J})=\ln \mathcal{Z}(\mathcal{J})$. For simplicity we assume that we have one replica field. The generating functional of one-particle irreducible correlations (vertex functions), $\Gamma[\overline{\phi}]$, is gotten by taking the Legendre transform of $\mathcal{W}(\mathcal{J})$ 

\begin{equation} 
\Gamma[\overline{\phi}]+\mathcal{W}(\mathcal{J})=\int d^{\,d}x\biggl(\mathcal{J}(x)\overline{\phi}(x)\biggr),
\end{equation}
\noindent where 

\beq
\overline{\phi}(x)=\left.\dfrac{\delta \mathcal{W}(\mathcal{J}) }{\delta \mathcal{J}}\right|_{\mathcal{J}=0}. 
\eeq

For sake of completeness we will discuss the one-loop renormalization of the corresponding theory. First, a vertex expansion for the effective action is given by

\begin{equation}
 \Gamma[\overline{\phi}]=\sum_{n=0}^{\infty}\frac{1}{n!}\int\prod_{i=1}^{n}{d^{\,d}x_{i}} \,\Gamma^{(n)}(x_{1},...,x_{n})\overline{\phi}(x_{1})...\overline{\phi}{(x_{n})},
\end{equation}  
where the expansion coefficients $\Gamma^{(n)}$ correspond to the one-particle irreducible (1PI) proper vertex. Writing the effective action in powers of momentum around the point where all 
external momenta vanish, we have

\begin{equation} 
\Gamma[\overline{\phi}]=\int\,d^{d}x \,V(\overline{\phi})+....
\end{equation}  
The term $V(\overline{\phi})$ is called the effective potential which takes into account the fluctuation in the model. Let us define the Fourier transform of the one-particle irreducible (1PI) proper vertex. We get

\begin{align}
 &\Gamma^{(n)}(x_{1},...,x_{n})=\frac{1}{(2\pi)^{n}}\int\prod_{i=1}^{n}{d^{\,d}k_{i}}(2\pi)^{d}\times \nonumber \\
 &\delta(k_{1}+...+k_{n})e^{i(k_{1}x_{1}+...+k_{n}x_{n})}\tilde{\Gamma}^{(n)}(k_{1},...,k_{n}).
\end{align} 

Now, we assume that the field $\overline{\phi}(x)=\phi$, is uniform. This condition is similar to the diluted instanton approximation. In this case, we 
can write 
\begin{equation} 
\Gamma[\phi]=\int d^{\,d}x\sum_{n=1}^{\infty}\frac{1}{n!}\biggl[\tilde{\Gamma}^{(n)}(0,...,0)\phi^{n}+...\biggr]
\end{equation}
\noindent The effective potential can be written as 

\begin{equation}
 V(\phi)=\sum_{n}\frac{1}{n!}\tilde{\Gamma}^{(n)}(0,...,0)\phi^{n}.
\end{equation} 
From above discussion it is possible to write the effective potential for each replica field in the leading replica partition function as $V(\phi)=V_{1}(\phi)+V_{2}(\phi)$, where 

\begin{align}
 V_{1}(\phi)=&\frac{1}{2}(m_{0}^{2}+\delta m^{2}_{0})\phi^{2}+\frac{1}{4}(g_{0}+\delta g_{0})\phi^{4} \nonumber \\
             &+\frac{1}{4}(\rho_{0}+\delta \rho_{0})\phi^{6} 
\end{align}  

\noindent $\delta m^{2}_{0}$, $\delta g_{0}$ and $\delta \rho_{0}$ are the counterterms that have to be introduced to remove divergent terms, and 

\begin{equation}
 V_{2}(\phi)=\frac{1}{2}\int\frac{d^{\,d}p}{(2\pi)^{d}}\ln\biggl[1+\frac{1}{p^{2}+m_{0}^{2}}(3g_{0}\phi^{2}+5\rho_{0}\phi^{4})\biggr].
\end{equation}
The calculation that we are presenting here is taking account the corrections due to the fluctuations around the saddle-point of each replica partition function. To proceed, we are interested to 
implement the one-loop renormalization in each replica field theory. The contribution to the effective potential given by $V_{2}(\phi)$ can be written as

\begin{equation}
 V_{2}(\phi)=\sum_{s=1}^{\infty}\frac{(-1)^{s+1}}{2s}\bigl(3g_{0}\phi^{2}+5\rho_{0}\phi^{4}\bigr)^{s}I(s,d),
\end{equation}
where $I(s,d)$ is given by

\begin{align}
 I(s,d)=&\int\frac{d^{\,d}q}{(2\pi)^{d}}\frac{1}{(q^{2}+m_{0}^{2})^{s}} \nonumber \\
       =&\frac{1}{(2\sqrt{\pi})^{d}}\frac{\Gamma(s-\frac{d}{2})}{\Gamma(s)}\bigl(m_{0}^{2}\bigr)^{\frac{d}{2}-s}.
\end{align}

At this point, let us use an analytic regularization procedure that has been used in field theory \cite{bol} and also to obtain the renormalized vacuum energy of a quantum field in the presence of boundaries \cite{nami1,nami2,nami3}. Using a well-known result that in the neighborhood of the pole $z=-n$ $(n=0,1,2,...)$ and for $\varepsilon\rightarrow 0$, the Gamma function has the representation 

\begin{equation}
 \Gamma(-n+\varepsilon)=\frac{(-1)^{n}}{n!}\left[\frac{1}{\varepsilon}+\psi(n+1)\right],
\end{equation}  
where $\psi(n+1)$, the digamma function, is the regular part in the neighborhood of the pole, and using the renormalization conditions which are given by

\begin{align}
&\frac{d^{\,2}}{d\phi^{\,2}}V(\phi)|_{\phi=0}=m_{R}^{2}, \nonumber \\
&\frac{d^{\,4}}{d\phi^{\,4}}V(\phi)|_{\phi=0}=g_{R},     \\
&\frac{d^{\,6}}{d\phi^{\,6}}V(\phi)|_{\phi=0}=\rho_{R},  \nonumber
\end{align} 
we obtain renormalized physical quantities. Note that the normalization conditions are choosen in the metastable vacuum state. It is possible to choose another normalization condition, as 
for example in the true minimum of the effective potential. We would like to stress that all the renormalization conditions are equivalent after one establish the correspondence between them \cite{colemanweinberg}. 
For the sake of simplicity, we consider the case where $d=4$:

\begin{align}
m_{R}^{2}&=m_{0}^{2}\left[1-\frac{3g_{0}\psi(2)}{16\pi^{2}}\right],\\
g_{R}&=6g_{0}+\frac{9\psi(1)}{4\pi^{2}}g_{0}^{2}-\frac{15\psi(2)}{4\pi^{2}}\rho_{0}m_{0}^{2},\\
\rho_{R}&=\rho_{0}\left[120-\frac{675\psi(1)}{2\pi^{2}}g_{0}\right]+\frac{369}{4\pi^{2}}\frac{g_{0}^{3}}{m_{0}^{2}},
\end{align} 

\noindent where $\psi(1)=-\gamma$ and $\psi(2)=-\gamma+1$.

\section{Conclusions}\label{sec:conclusions}

A fundamental question is the following: how the statistical mechanics mathematical formalism of disordered systems differs from that homogeneous systems? In homogeneous systems the study of the low-temperature phase can be simplified making use of many spatial symmetries that such systems have. In principle, in quenched disordered systems these symmetries are absent. The first step to recover at least the translational symmetry in disordered systems is averaging over the quenched disorder. Here, in order  to partially answer the above question we would like to discuss a few disordered models that has been investigated in the lattice and also using  statistical field theory defined in the continuum. 

One of the simplest model of spin-glass is the Edwards-Anderson model. The spin-glass phase of this model is characterized by the absence of orientational localized magnetic moments ordering in space at low temperatures. This indicates that the system does not have a unique ground state. These multiple vacuum states appear in a infinite range spin model, the Sherrington-Kirkpatrick model.
In this model a replica symmetry breaking mechanism was introduced in order to prevent the emergence of unphysical results, i.e., a negative entropy at low temperatures, which would arise with the assumption of a replica-symmetric solution in such model.  In the replica symmetry breaking scheme for this fully-connected model the mean-field approximation predicts a unusual structure of the free energy: a existence of a  multivalley structure in the the free energy landscape. Studying a statistical field theory defined in the continuum, some authors using the replica method with replica symmetry breaking mechanism, discussed the spin glass-like behavior and the possibility of a existence of infinitely many ground states in the random temperature Landau-Ginzburg model.

One of the motivations of this paper is to emphasize the differences and similarities between the DZFM and the conventional replica method, discussing a disordered $\lambda\varphi^{4}+\rho\varphi^{6}$ Landau-Ginzburg model defined in a $d$-dimensional space.  First we adopt the standard procedure averaging the disorder dependent free energy using the DZFM. We show that the dominant contribution to the average free energy of this system is written as a series of the replica partition functions of the model.  In a generic replica partition function, the structure of the replica space is investigated using the saddle-point equations. In each replica partition function we must impose the replica symmetry ansatz. We prove that the average free energy represents a system with multiple ground states with different order parameters. This situation is quite  similar to the one obtained in a fully connected mean-field model in a replica symmetry breaking scenario. For low temperatures we show also the existence of metastable equilibrium states for some replica fields. In the low temperature regime, one way to proceed is to consider the possibility that the series representation for the average free energy describes inhomogeneous domains, i.e., macroscopic regions in a sample $\Omega^{(k)}$ with at least one proper characteristic order parameter $\varphi^{(k)}_{0}$.
 
Finally we discuss the leading term in the series representation for the average free energy. This leading term of this series expansion is a large-$N$ Euclidean replica field theory. In this leading order replica partition function, the one-loop renormalization of this model is performed. It is important to point out that it is possible to go beyond the one-loop approximation using the composite field operator formalism \cite{jackiw,gino1,gino2, gino3,tarjus,periodic}, where an infinite of leading diagrams is summed. The technique deal with the effective action formalism for composite operators. One must consider a generalization of the effective action where the scalar field is coupled linearly and quadratically to sources. This generalization is under investigation by the authors.

\section{Acknowlegements} This work was partially suported by Conselho Nacional de Desenvolvimento Cient\'{\i}fico e Tecnol\'{o}gico - CNPq, 305894/2009-9 (G.K.), 303436/2015-8 (N.F.S.) and Funda\c{c}\~{a}o de Amparo \`{a} Pesquisa do Estado de S\~{a}o Paulo - FAPESP, 2013/01907-0 (G.K).

\end{document}